\documentclass[aps,prl,preprintnumbers,twocolumn,groupedaddress]{revtex4}

\usepackage{graphicx}

\begin{document}

\preprint{UMD-PP-05-026}

\title{Suppressing Proton Decay in the Minimal SO(10) Model}


\author{Bhaskar Dutta$^*$}
\author{Yukihiro Mimura$^*$}
\author{R.N. Mohapatra$^\dagger$}
\affiliation{$^*$Department of Physics, University of Regina, Regina, Saskatchewan S4S 0A2, Canada \\
             $^\dagger$Department of Physics, University of Maryland, College Park, MD 20742, USA}


\date{\today}

\begin{abstract}
We show that in a class of minimal supersymmetric SO(10) models
which have been found to be quite successful in predicting
neutrino mixings, all proton decay modes can be suppressed by  a
particular choice of Yukawa textures. This suppression works for
contributions from both left and right operators for nucleon
decay and for arbitrary $\tan\beta$. The required texture not only
fits all lepton and quark masses as well as CKM parameters  but it
also predicts neutrino mixing parameter $U_{e3}$ and Dirac CP
phase $\sin|\delta_{MNS}|$ to be 0.07-0.09 and 0.3-0.7
respectively. We also discuss the relation between the GUT
symmetry breaking parameters for the origin of these textures.
\end{abstract}


\maketitle


\section{Introduction}
Recent experiments in neutrino physics which have established that
neutrino masses are in the sub-eV range have provided a new reason
for taking supersymmetric grand unified theories (SUSY GUTs)
seriously. This has to do with using the seesaw mechanism to
understand the extreme smallness of neutrino masses compared to
charged fermion masses\cite{seesaw}. The seesaw mechanism
consists of extending the standard model by introducing three
right-handed neutrinos with large Majorana masses $M_R$. Simple
arguments based on the atmospheric oscillation data then tell us
that at least one $M_R$ is
 close to the conventional GUT scale $M_U\sim 10^{16}$ GeV,
  perhaps suggesting that the seesaw scale,
$M_R$ and GUT scale are one and the same.

 The minimal GUT that
unifies the right-handed neutrinos with the other fermions of the
standard model and leads to the seesaw mechanism is the SO(10)
model where all quarks and leptons are unified into one {\bf 16}
dimensional spinor multiplet. This raises the possibility that
the variety of masses and mixings of quarks and leptons can be
understood in terms of a smaller number of parameters than in the
standard model with three right-handed neutrinos. In fact,
various recent works in the minimal SO(10) models with a single
{\bf 10} and a single $\overline{\bf 126}$ Higgs
multiplets\cite{babu,other,Bajc:2002iw,Goh:2003sy,Dutta:2004wv}
have substantiated this point of view. Unlike the generic seesaw
models based on the extension of the standard model as well as
right-handed neutrino extended SU(5) model, the above SO(10)
models have fewer parameters and lead to predictions for neutrino
mixings and phases in their minimal version without any extra
symmetry assumptions. Their predictions for solar and atmospheric
mixing angles are in agreement with present
observations\cite{Goh:2003sy,Dutta:2004wv} and that for $U_{e3}$
is not far below the present upper limits, making the model
testable in planned experiments\cite{reactor}. Measurement of
$U_{e3}$ will therefore provide a crucial test of these models.

 It has further been shown that the simplest way to accommodate CKM CP
violation in these models is to include an additional Higgs field
belonging to the {\bf 120} dimensional
representation\cite{Dutta:2004hp}. The model with {\bf 120} still
remains predictive in the neutrino sector despite a small
increase in the number of
parameters\cite{Dutta:2004hp,Bertolini:2004eq} and also leads to
a solution to the SUSY CP problem.

Since the quarks and leptons are unified in a GUT, it
 predicts that proton is no longer stable and therefore
 proton decay becomes another test of any GUT such as SO(10).
The present experimental bounds on the proton lifetime are in
fact known to severely constrain some GUT
models\cite{Hisano:2000dg,Goto:1998qg} and one must therefore
make sure that the above class of SO(10) models are in agreement
with present experimental results.

In most generic SUSY GUTs, the dimension five operators induced by
colored Higgsino\cite{Sakai:1981pk} provide the dominant
contribution to the proton decay amplitude rather than the
dimension six operators induced by gauge bosons. Since the
dimension five operators arise from diagrams involving Yukawa
couplings, predictions for proton lifetime get related  to
fermion mass textures. For example, in simple GUT theories such
as minimal SU(5),  where by minimal we
mean that only the most general renormalizable terms are included in the
superpotential, the up- and down-type quark mass matrices are
proportional to the Yukawa matrices, $Y_u$ and $Y_d$ respectively. In this
case the proton decay rate, $\tau^{-1}_p$ is directly proportional to
$|Y_{u,ij}Y_{d,kl}|^2$ implying that $\tau_p$ cannot be
arbitrarily adjusted. That is why the minimal SU(5) theory is now
ruled out by proton decay results\cite{Goto:1998qg}.

One might argue that the minimal SU(5) model is
anyway not realistic
 since it predicts wrong relations between fermion masses e.g.
$m_s\,=\,m_\mu$ and $m_d\,=\,m_e$. This fermion mass problem is
however easily remedied in the class of minimal SO(10)
models\cite{babu} with a single {\bf 10} and $\overline{\bf 126}$
field mentioned above.  In view of the fact that it is also quite
predictive in the neutrino
sector\cite{babu,Bajc:2002iw,Goh:2003sy,Dutta:2004wv}, it is
tempting to consider this as the minimal GUT comparable to what
SU(5) GUT was in the 1980's. It is then important to look for its
predictions for proton decay. To be sure, this model like minimal
SU(5) (and the standard model) does not explain the origin of the
Yukawa couplings; however if this model is confirmed by
experiments, understanding the Yukawa sector will be the next
order of business.

Unlike the SU(5) model,  the $\Delta B =1$ interactions in
generic SO(10) models\cite{babu1} involve more GUT scale symmetry
breaking parameters than just the color triplet Higgsino mass;
therefore the situation for proton decay here is  less
restrictive. In particular, in the minimal SO(10) models of the
type discussed in Ref.\cite{Goh:2003sy}, there are four free
parameters\cite{Goh:2003nv} whereas there are about 16 decay
modes which have lower bounds on their partial life times. It was
shown through a numerical
analysis\cite{Goh:2003nv,Fukuyama:2004pb}
 (without including $RRRR$ operators) that there exists a very
small region in these parameter spaces for $LLLL$ operators, where all
the present experimental constraints are satisfied for lower
$\tan\beta$.

In this paper, we study the proton decay constraints on SO(10) models with
$\bf 10+\overline{126}+120$ Higgs fields.
We show that present proton decay constraints combined with
fermion masses and mixings imply
a specific relation among SO(10) breaking vacuum expectation values (VEVs) and
 a very specific form for the Yukawa textures.
Roughly, they imply that the proton decay operators are proportional to the
product of two up-type quark Yukawa couplings 
i.e. $\!Y_{u,ij}Y_{u,kl}$ instead of $Y_d Y_u$ as in SU(5). As a consequence,
proton decay is not only suppressed for the left-handed quark
contributions but also for right-handed ones and furthermore, the
suppression works also for large $\tan\beta$ values as well which
makes this model easily distinguishable from other simple GUT
models. In addition, it leads to definite predictions for the neutrino
mixing
parameters $U_{e3}$ and the Dirac phase $\delta_{MNS}$. We also show the
essential role played by the {\bf 120} in the suppression of  proton decay
when all operators are taken into account.
 These are the new results of this paper, which have important
implications for the viability of the minimal SO(10) model for neutrino
masses with CKM CP violation.

\section{Fermion mass and proton decay in minimal SO(10)}
We start by introducing the Yukawa interactions and the contents
of Higgs fields in the SO(10) model.
The Yukawa superpotential involves the couplings of $\bf 16$-dimensional
matter spinors $\psi_i$ ($i$ denotes a generation index) with
$\bf 10$ ($H$), $\overline{\bf 126}$ ($\overline\Delta$), and
$\bf 120$ ($D$) dimensional Higgs fields:
\begin{equation}
W_Y = \frac12 h_{ij} \psi_i \psi_j H + \frac12 f_{ij} \psi_i \psi_j
\overline\Delta
+ \frac12 h^\prime_{ij} \psi_i \psi_j D.
\end{equation}
The Yukawa couplings, $h$ and $f$ are symmetric matrices, whereas $h^\prime$
is an anti-symmetric matrix due to SO(10) symmetry.
One $\bf 126$ Higgs multiplet $\Delta$ is also introduced
as a vector-like pair of $\overline\Delta$ whose VEV
reduces the rank of SO(10) group. This helps to keep supersymmetry
unbroken down to the weak scale.
In order to break SO(10) symmetry down to the standard model,
we employ one $\bf 210$ Higgs field ($\Phi$)\cite{aul}
which also contains a pair of Higgs doublets ($\Phi_u,\,\Phi_d$).
Altogether, we have six pairs of Higgs doublets:
$\varphi_d = (H^{10}_d, D^{1}_d, D^{2}_d,
\overline\Delta_d, \Delta_d, \Phi_d)$,
$\varphi_u = (H_{u}^{10}, D_{u}^1, D_{u}^2,
\Delta_u, \overline\Delta_u,  \Phi_{u})$,
where superscripts $1$, $2$ of $D_{u,d}$ stand for SU(4) singlet and adjoint
pieces under the $G_{422}=\ $SU(4)$\times $SU(2)$\times
$SU(2) decomposition.
The mass term of the Higgs doublets is given as
$(\varphi_d)_a (M_D)_{ab} (\varphi_u)_b$,
and the expression of the matrix $M_D$ is given in Ref.\cite{Fukuyama:2004ps}.
The mass matrix of the Higgs doublets is diagonalized by unitary matrices
$U$ and $V$:
$U M_D V^{\rm T} = M_D^{\rm diag}$.
We assume that $(M_D^{\rm diag})_{11} = \mu$, where $\mu$ is a Higgsino mass
in the MSSM and the mass scale is much smaller than the GUT scale.
Since we concentrate on the structure of Yukawa couplings,
we do not need to specify the dynamical reason of the mass hierarchy in
this letter,
but we just require a fine-tuning such as in the case of minimal
SU(5) model\cite{Hisano:2000dg}.
The MSSM Higgs doublets are given as linear combinations:
$H_d = U_{1a}^* (\varphi_d)_a$, $H_u = V_{1a}^* (\varphi_u)_a$.

We use ``Y diagonal basis" (or SU(5) basis) to
describe the standard
model decomposition of the SO(10) representation\cite{Fukuyama:2004ps,Aulakh:2003kg}.
The expression of the Yukawa interaction under the $G_{422}$ decomposition
can be derived from  Ref.\cite{Aulakh:2002zr}.
The decomposed Yukawa interactions which give fermion masses are written as
\begin{eqnarray}
\!\!\!\!&\!\!\!\!&\!\!\!\!
W_Y^{\rm doub.} = h H_d^{10} (q d^c + \ell e^c) + h H_u^{10} (q u^c + \ell \nu^c) \\
&&+          \frac1{\sqrt3}f \overline\Delta_d (q d^c -3 \ell e^c)
         + \frac1{\sqrt3}f \overline\Delta_u (q u^c - 3 \ell \nu^c)  \nonumber \\
&&+   \     h^\prime D_d^1 (q d^c + \ell e^c) + h^\prime D_u ^1 (q u^c + \ell \nu^c) \nonumber\\
&&+          \frac1{\sqrt3}h^\prime D_d^2 (q d^c -3 \ell e^c)
         - \frac1{\sqrt3}h^\prime D_u ^2 (q u^c -3 \ell \nu^c), \nonumber
\end{eqnarray}
where $q,u^c,d^c,\ell,e^c,\nu^c$ are the quark and lepton fields for the standard model,
which are all unified into one spinor representation of SO(10).
%
%
%
We obtain the Yukawa coupling matrices for fermions as
\begin{eqnarray}
Y_u &=& \bar h + r_2\bar f + r_3 \bar h^\prime, \label{Y_u} \\
Y_d &=& r_1(\bar h + \bar f + \bar h^\prime), \\
Y_e &=& r_1(\bar h - 3 \bar f + c_e\bar h^\prime), \label{Y_e}\\
Y_\nu &=& \bar h - 3 r_2\bar  f + c_\nu \bar h^\prime, \label{Y_nu}
\end{eqnarray}
where the subscripts $u,d,e,\nu$ denotes for up-type quark,
down-type quark, charged-lepton, and Dirac neutrino Yukawa couplings,
respectively and
\begin{eqnarray}
\bar h &\!\!=&\!\! V_{11} h,\ r_1 = U_{11}/V_{11},\
r_2 = r_1 V_{15}/U_{14}, \\
r_3 &\!\!=&\!\! r_1 (V_{12} -V_{13}/\sqrt3)/(U_{12} + U_{13}/\sqrt3), \\
\bar f &\!\!=&\!\! U_{14}/(\sqrt3\, r_1) f,\
\bar h^\prime = (U_{12} + U_{13}/\sqrt3)/r_1 h^\prime,\\
c_e &\!\!=&\!\! (U_{12} -\sqrt3 U_{13})/(U_{12} + U_{13}/\sqrt3),\\
c_\nu &\!\!=&\!\! r_1 (V_{12} +\sqrt3 V_{13})/(U_{12} +
U_{13}/\sqrt3).
\end{eqnarray}
The Majorana mass matrices for both left- and right-handed neutrinos
are proportional to the coupling $f$.
In this letter, we will be using type II seesaw\cite{seesaw2}.

Next we consider dimension five operators induced by Higgs triplets.
The dimension five operators ($LLLL$ and $RRRR$ operators),
\begin{equation}
-W_5 =\frac12 C_L^{ijkl} q_k q_l q_i \ell_j + C_R^{ijkl} e_k^c u_l^c
u_i^c d_j^c ,
\end{equation}
are obtained by integrating out the triplet Higgs fields,
$\varphi_{\overline T}=(H_{\overline T},D_{\overline T},D_{\overline
T}^\prime,\overline
\Delta_{\overline T},\Delta_{\overline T},\Delta_{\overline T}^\prime,
\Phi_{\overline T})$
and
$\varphi_T=(H_{T},D_{T},D_{T}^\prime,\Delta_{T},\overline\Delta_{T},
\overline\Delta_{T}^\prime,  \Phi_{T})$.
The quantum numbers under SU(3)$\times$SU(2)$\times$U(1)$_Y$
of the field $\varphi_T$ is $({\bf 3},{\bf 1},-1/3)$.
In the expression of $\varphi_T$, the fields with  `$^\prime$'
are decuplet, and the others are sextet under SU(4) decomposition.
The $RRRR$ operator, $C_R$, is also generated by other colored triplet,
$\varphi_{\overline C} = (D_{\overline C}, \Delta_{\overline C})$
and $\varphi_C = (D_C, \overline\Delta_C)$, where the quantum number
of $\varphi_C$ is $({\bf 3}, {\bf 1}, -4/3)$.
The mass term of the Higgs triplets are
given as
$(\varphi_{\overline T})_a (M_T)_{ab} (\varphi_T)_b +
(\varphi_{\overline C})_a (M_C)_{ab} (\varphi_C)_b$.
The mass matrices, $M_T$ and $M_C$, are 7$\times$7 and 2$\times$2 matrices
respectively, and their explicit forms are given in the
literature\cite{Fukuyama:2004ps}.
The Yukawa couplings which cause proton decay are written as
\begin{eqnarray}
&\!\!\!&\!\!\! W_Y^{\rm trip.} = h H_{\overline T}\ (q \ell + u^c d^c)
+ h H_{T}\ (\frac12 qq + e^c u^c )\nonumber \\
&\!\!+&\!\! f \overline\Delta_{\overline T}\, (q \ell -  u^c d^c)
+ f \overline\Delta_{T}\, (\frac12 qq -  e^c u^c ) \nonumber
+{\sqrt2} f \overline\Delta_{T}^\prime\ e^c u^c
\\
&\!\!+&\!\! {\sqrt2} h^\prime D_{\overline T}\ u^c d^c
+ {\sqrt2} h^\prime D_{\overline T}^\prime\ q \ell \\
&\!\!&\!\! \quad \quad {-\sqrt2} h^\prime D_{T}\ e^c u^c
+ {\sqrt2} h^\prime D_{T}^\prime\ e^cu^c  \nonumber\\
 &\!\!+&\!\!  2 f \overline\Delta_{C} \, d^c e^c + 2 h^\prime
D_{\overline C} \ u^c u^c
+2 h^\prime D_{C} \ d^c e^c. \nonumber
\end{eqnarray}
The dimension five operators are
written by the Yukawa couplings $h$, $f$ and $h^\prime$ as follow:
\begin{eqnarray}
&&C_L^{ijkl} = c \, h_{ij}h_{kl}+ x_1 f_{ij}f_{kl} +
x_2 h_{ij}f_{kl} + x_3 f_{ij}h_{kl} \nonumber \\
&& +
x_4 h^\prime_{ij}h_{kl} + x_5 h^\prime_{ij}f_{kl} , \label{LLLL} \\
&& C_R^{ijkl}
= c \, h_{ij}h_{kl}+ y_1 f_{ij} f_{kl} + y_2 h_{ij}f_{kl} + y_3
f_{ij}h_{kl} \nonumber
\\
&&+ y_4 h^\prime_{ij}h_{kl}  + y_5 h^\prime_{ij}f_{kl}
 + y_6  h_{ij}h^\prime_{kl} + y_7  f_{ij}h^\prime_{kl} + y_8
h^\prime_{ij} h^\prime_{kl} \nonumber
\\
&&+ y_9 h^\prime_{il} f_{jk} + y_{10} h^\prime_{il} h^\prime_{jk}. \label{RRRR}
\end{eqnarray}
The common coefficient $c$ is given as $c=(M_T^{-1})_{11}$.
We obtain the other coefficients as the following:
\begin{eqnarray}
&&(x_1,x_2,x_3,x_4,x_5)  \nonumber \\
&=& ( M^{-1}_{54}, M^{-1}_{51}, M^{-1}_{14},
{\sqrt2} M^{-1}_{13}, {\sqrt2} M^{-1}_{53}) ,  \\
&& (y_1,y_2,y_3,y_4,y_5,y_6,y_7,y_8) \nonumber\\
&=& ( M^{-1}_{54} {-\sqrt2}M^{-1}_{64},
{-}M^{-1}_{51} + {\sqrt2}M^{-1}_{61},
{-}M^{-1}_{14},  \label{RRRR2}\\
&& \sqrt2 M^{-1}_{12},
{-\sqrt{2}}M^{-1}_{52}+{2}M^{-1}_{62},
{\sqrt2}({-}M^{-1}_{21} + M^{-1}_{31}), \nonumber \\
&& \sqrt2 ( M^{-1}_{24} {-} M^{-1}_{34}),
2 ({-}M^{-1}_{22}+M^{-1}_{32})
), \nonumber
\end{eqnarray}
where subscript $T$ of the matrix $M_T$ is omitted.
It is important to note that $y_3 = - x_3$,
namely their signatures are opposite between $LLLL$ and $RRRR$ operator.
This is derived from the fact that $H_T$ and $\overline\Delta_T$ have
opposite D-parity.
The coefficients $y_9$, $y_{10}$ are generated by
$\varphi_C$, $\varphi_{\overline C}$,
and $y_9 = 4 (M_C^{-1})_{21}$, $y_{10} = 4 (M_C^{-1})_{11}$.

The Eqs.(\ref{LLLL}-\ref{RRRR2}) are explicit forms
of the dimension five operators, but we present here  more convenient
forms to describe
proton decay.
 The Higgs triplet mass matrix $M_T$ is diagonalized by two unitary
matrices, $X$ and $Y$,
as $X M_T Y^{\rm T} = {\rm diag} (M_1, M_2, \cdots, M_7)$.
Then we obtain the useful formula for dimension five operators
\begin{eqnarray}
&\!\!\!&\!\!\! C_L^{ijkl}
= \sum_a \frac1{M_a}(X_{a1} h + X_{a4} f + \sqrt2 X_{a3} h^\prime)_{ij}
\label{LLLL3}\\
&& \qquad \qquad \qquad \qquad \qquad  \times(Y_{a1} h + Y_{a5} f)_{kl}
,\nonumber \\
&\!\!\!&\!\!\! C_R^{ijkl}
= \sum_a \frac1{M_a} (X_{a1} h - X_{a4} f + \sqrt2 X_{a2} h^\prime)_{ij}
\label{RRRR3}\\
&& \times (Y_{a1} h - (Y_{a5}- \sqrt2 Y_{a6}) f + \sqrt2
(Y_{a3}-Y_{a2}) h^\prime )_{kl} \nonumber \\
&&+(y_9, y_{10}\ {\rm terms}) \nonumber.
\end{eqnarray}

One can make consistency checks to verify the formula by
considering specific vacua of the theory. For example, in the
SU(5) limit, when only one of the colored triplets is much lighter
than the others, i.e. $M_1 \ll M_a$ $(a\neq 1)$, we can obtain the
following relations for the diagonalizing matrices from the
explicit form of the Higgs mass matrices with the SU(5) vacua in
the Ref.\cite{Fukuyama:2004ps,Aulakh:2003kg}:
\begin{eqnarray}
U_{11}=X_{11}, \ V_{11}&\!=&\!Y_{11}, \ U_{14}=X_{14}=0
,\label{su5-1} \\
V_{15}: Y_{15} : Y_{16} &\!=&\! \sqrt3 : 1 : \sqrt2 , \label{su5-2} \\
U_{12} : U_{13} : X_{12} : X_{13} &\!=&\! V_{12} : V_{13} : Y_{12}
: Y_{13} \label{su5-3} \\
&\!=&\!1 : \sqrt3 : \sqrt2 : \sqrt2,\label{su5-4}
\end{eqnarray}
As a result,
$r_2 \rightarrow \infty$ with $\bar f \rightarrow 0$,
\begin{equation}
 r_3 = 0, \ c_e = -1 \label{su5-5}
\end{equation}
for the Yukawa matrices in Eqs.(\ref{Y_u}-\ref{Y_nu})
and thus, as expected, we get the SU(5) relations, 
$Y_u = Y_u^{\rm T}$,
$Y_d = Y_e^{\rm T}$,
and the dimension five proton decay operators can be written in terms of
the Yukawa couplings as
\begin{eqnarray}
C_L^{ijkl} \simeq C_R^{ijkl} \simeq (Y_d)_{ij} (Y_u)_{kl}/M_1. \label{su5}
\end{eqnarray}

One can also obtain the flipped-SU(5) limit\cite{f5} in a
similar manner and find the relations, $Y_u = Y_\nu^{\rm T}$,
$Y_d = Y_d^{\rm T}$,
and
\begin{eqnarray}
C_L^{ijkl} \simeq (Y_u)_{ji} (Y_d)_{kl}/M_1, \
C_R^{ijkl} \simeq (Y_u)_{ij} (Y_e)_{lk}/M_1. \label{flipped}
\end{eqnarray}
One can also consider other typical vacua such as the one corresponding to
$G_{422}$.
On this vacuum, we have 
$U_{13}=V_{13}=U_{14}=V_{15}=0$
and all four fermion Yukawa couplings are unified as
$Y_{u,d,e,\nu} = \bar h + \bar h^\prime$, which is not viable
for phenomenology. In any case,
since the doublet and triplet Higgs fields are not unified in one
multiplet on this vacuum,
proton decay operator cannot be written in terms of fermion Yukawa couplings
contrary to the SU(5) and flipped-SU(5) limit.
The operators $C_{L,R}^{ijkl}$ in the $G_{422}$ limit
are symmetric under the interchange of indices $ij$ and $kl$
in addition to the symmetry under the exchange of individual indices
$i$ and $j$.
One can write down the dimension five operators as
\begin{eqnarray}
C_L \propto (h+zf)(h+zf), \ C_R \propto (h-zf)(h-zf),
\end{eqnarray}
with $z= X_{14}/X_{11}$.
It is important to stress that there are relatively opposite signatures
for the $f$ contributions between
$C_L$ and $C_R$. This is due to D-parity, which we have already mentioned.

\section{Suppression of proton decay}

Let us now investigate the conditions required to suppress the
proton decay rate in the minimal SO(10) model described above. We
first note that the four-fermion proton decay operators are
produced by gaugino and Higgsino dressing of the dimension five
operators. The four-fermion operators of $LLLL$ type get dominant
contribution from wino exchange and therefore retain the same
flavor structure as that of the original dimension five
supersymmetric operator (we have  assumed universality of scalar
masses  for suppressing FCNC processes). However, as far as the
$RRRR$ operators are concerned, they receive contributions only
from Higgsino exchange which can involve the top quark and $\tau$
lepton Yukawa couplings. For instance in the case of SU(5) model,
the contribution of the $RRRR$ operator to the decay mode
$p\rightarrow K \bar\nu_\tau$ gets so enhanced because of this
that it exceeds the current experimental bound for any
$\tan\beta$ as long as stop mass is less than around 1
TeV\cite{Goto:1998qg}.

Since the original Yukawa couplings $h$, $f$, and $h^\prime$ are
 functions of fermion Yukawa couplings $Y_u$, $Y_d$ and $Y_e$ via
Eqs.(\ref{Y_u}-\ref{Y_e}),
their textures are roughly determined from the experimental inputs of
quark and lepton masses and mixings. For instance,
 to fit the strange quark
and muon masses, bottom-tau unification, relations among CKM mixings
and also large mixings of neutrinos,
the coupling $f$ is almost determined to have the form
$\bar f \sim  \left(
               \begin{array}{ccc}
                \lambda^2 & \lambda & \lambda \\
                \lambda & 1 & 1 \\
                \lambda & 1 & 1
               \end{array}
         \right) {m_s}/{m_b}$, where $\lambda \sim 0.2$.
A naive implication of this is that since up-type quark masses
are more hierarchical than down-type ones (i.e. $m_u/m_t \ll
m_d/m_b$, $m_c/m_t \ll m_s/m_b$), the expression for the up-type
Yukawa matrix, $Y_u$, in Eq.(\ref{Y_u}) requires the following two typical choices:
(a) there is cancellation among $h$, $f$, and $h^\prime$, or
(b) $h$ itself has a hierarchical form similary to the up-type quark masses.
The first choice (a) is the case where [1,2] block of $h_{ij}$ is not far smaller
than $f_{ij}$, but $r_{2,3}$ are chosen to be certain values
to make $m_u$, $m_c$ are hierarchically small.
The second choice (b) corresponds to the case where $r_{2,3}\sim 0$.
The second choice appears to be required to suppress the
proton decay. Let us discuss the reason. 

In order to suppress
the  decay rate, we need small couplings for first and second
generations in the expressions in Eqs.(\ref{LLLL}-\ref{RRRR}).
Clearly this would also require a cancellation among $h$, $f$ and
$h^\prime$ if we take the first choice (a). 
Since in general the coefficients $r_2, r_3$ in up-type Yukawa matrix 
and $x_i$ and $y_i$ in proton decay operators are
unrelated, one must find a situation where both cancellations can
be achieved in a satisfactory manner so as to be consistent with
all data.
However, if we take the choice (a), the cancellation cannot happen naturally
due to the following reasons.


First, let us discuss the $\overline{\bf 126}$ Higgs contribution.
Since there is an opposite signature for one of the coefficients in the $LLLL$
and $RRRR$ operators, $y_3 = - x_3$, the
cancellation required to obtain small Yukawa coupling for $Y_u$ by
tuning $r_2 \bar f$ cannot simultaneously suppress both $LLLL$ and
$RRRR$ operators. Thus, in general, it is hard to suppress proton
decay rate by tuning $\overline{\bf 126}$ colored Higgs mixing
$X_{14}$.
%
%
%
%
%
%
%
Next let us see the contribution from $\bf 120$ Higgs field.
The coefficient $C_L^{ijkl}$ is symmetric in the
indices $kl$ due to SU(3)$\times$SU(2) contraction. Therefore
$h^\prime$ contribution is absent for the $q_k q_l$ part in
$LLLL$ operator, whereas it is of course present in the fermion
masses. Thus, if the cancellation in $Y_u$ happens by tuning $r_3
h^\prime$, such cancellation will not help in suppressing the
$LLLL$ operator. 

The above discussions lead to the fact that the proton decay rate cannot be 
suppressed in natural way if we take the choice~(a).
As we will see numerical studies later, we require a fine-tuning
to the level of 0.01\% to suppress all the proton and neutron
decay modes when we consider general parameter fitting in the choice (a).
The choice~(b) ($r_{2,3} \simeq 0$) is necessary
to achieve natural suppression of proton decay as a result.

We now show how the proton decay rate in the choice~(b)
is suppressed rather than the minimal SU(5).
%
%
If $r_{2,3} \simeq 0$,
the $LLLL$ dimension five operator can be a form as 
$C_L^{ijkl} \propto (Y_u + \gamma h^\prime)_{ij}(Y_u)_{kl}$ 
and in the $RRRR$ operator $C_R^{ijkl}$, 
$kl$ part is also related to $Y_u$. 
This will correspond to the case where $X_{14},Y_{15}
\sim 0$. We
will give an example later where this can happen but first we note
that for the case just noted, the $RRRR$ contribution to 
$p \rightarrow K \bar\nu_\tau$ mode is suppressed compared to the
usual SU(5) models for the entire range of tan$\beta$ up to $50$.
Here, the dominant contribution is proportional to $\lambda_u$ 
giving a suppression factor $\lambda_u/\lambda_d\sim 1/100$ for
$\tan\beta\sim 50$ compared to minimal SU(5) case. 
%
Similarly, since the $kl$ part of $C_L$ are also related to the
$Y_u$ instead of $Y_d$, the $LLLL$ contribution to the $p
\rightarrow K \bar\nu$ is also suppressed even for
$\tan\beta \sim 50$, compared to the SU(5) model (since
$\lambda_c/\lambda_s\sim1/5$). However as it turns out, these
suppressions are not enough. We need to specify the Yukawa texture for the
purpose as we see below.

Before describing the specific choice of Yukawa textures, 
let us show the numerical values for the proton decay amplitude
to confirm that $r_{2,3}\simeq 0$ is necessary to satisfy the current 
experimental bounds naturally.
Rewriting the proton decay amplitude
as $A = \alpha_2 \beta_p/(4\pi M_T m_{SUSY}) \tilde A$, we define
\begin{equation}
\tilde A = c\tilde A_{hh}+ x_1\tilde A_{ff}+x_2\tilde A_{hf}+
x_3\tilde A_{fh} +x_4\tilde A_{h^\prime h}+ \cdots.
\end{equation}
The coefficients $c$ and $x_i$ are given in Eq.(\ref{LLLL}).
The $RRRR$ contribution can be written in the same way.
To satisfy the current proton and neutron decay bounds, we need
$|\tilde A_{p\rightarrow K\bar\nu}| \alt 10^{-8}$,
$|\tilde A_{n\rightarrow \pi\bar\nu}| \alt 2 \cdot 10^{-8}$
and $|\tilde A_{n\rightarrow K\bar\nu}| \alt 5 \cdot 10^{-8}$
if the colored Higgsino mass is $2 \cdot 10^{16}$ GeV, and squark and
wino masses are
around 1 TeV and 250 GeV, respectively.
In order to satisfy those bounds naturally, we need $\tilde A_{hh} \alt 5
\cdot 10^{-8}$.
If $\tilde A_{hh} \agt 10^{-7}$,
we need to tune $x_i$ and $y_i$ for every decay mode to cancel $\tilde
A_{hh}$, which is unnatural.
(Further, assuming $c \rightarrow 0$ cannot make successful suppression
of the decay amplitude since in that assumption we also need
 to suppress $\tilde A_{ff}$ which can only be suppressed for  $\tan\beta
\alt 3$).
%
The $\tilde A_{hh}$ depends on the magnitudes of the elements from the
[1,2] block of $\bar h$ which is
determined from the fit to the up-type Yukawa as a function of $r_2$ and $r_3$.
In Table {\ref{tabr2r3}} we show the $\tilde A_{hh}$ component
in the proton decay amplitude for different values of
$r_3$ and $r_2$.
\begin{table}
\caption{\label{tabr2r3}Typical proton decay amplitude for different values of
$r_2$ and $r_3$.
To satisfy current experimental bound naturally, $|\tilde A_{hh}| \alt 10^{-8}$ is
needed.}
 \begin{ruledtabular}
 \begin{tabular}{|c|cc|}
 $r_2$, $r_3$&  $\tilde A_{hh}$ &\\\hline
 $r_2=r_3$= 0 & $(4.2+0.083 i) \times 10^{-7}$ &\\
 $r_2$= 0, $r_3$= 1 & $(1.4-0.17 i) \times 10^{-4}$ &\\
 $r_2$= 0.5, $r_3$= 0 & $(7.3+2.6 i)\times 10^{-5}$ &\\
 \end{tabular}
 \end{ruledtabular}
 \end{table}
The $\tilde A_{hh}$ is calculated in the basis where
$Y_u={\rm diag}(\lambda_u,\lambda_c,\lambda_t)$. As we can see from the table, that
$\tilde A_{hh}$ can be very large (for large $r_{2,3}$) and this
requires very high level of fine-tuning for all the decay
modes (including charged leptonic decay) to suppress nucleon
decay as we noted. The smallness of $r_2 $ and $r_3$ 
 is required to suppress $\tilde A_{hh}$. 
However, even in the case $r_2=r_3=0$, 
the $\tilde A_{hh}$ amplitude is not enough suppressed
($\tilde A_{hh}\sim 10^{-7}$) and we
still need fine-tuning among the coefficients to satisfy the data.
This is mainly because $\lambda_{u,c}$ are not enough small to satisfy
$\tilde A_{hh} \alt 10^{-8}$. 

We now present the specific choice of Yukawa texture to suppress 
the proton decay rate further.
The texture is given as
$\bar h \simeq $ diag$(0,0,O(1))$, so that up and
charm quark masses arise from contributions of 
$f$ and $h'$:
\begin{equation}
\bar f \simeq \left(\begin{array}{ccc}
\sim 0 & \sim 0 & \lambda^3 \\
\sim 0 & \lambda^2 & \lambda^2 \\
\lambda^3 & \lambda^2 & \lambda^2
\end{array}
\right) , \
\bar h^\prime \simeq i \left(\begin{array}{ccc}
0 & \lambda^3 & \lambda^3 \\
-\lambda^3 & 0 & \lambda^2 \\
-\lambda^3 & -\lambda^2 & 0\end{array}
\right),
\end{equation}
where $\lambda \sim 0.2$. In this texture, $r_3 = 0$,
and $r_2$ is fixed as $r_2 m_s/m_b \simeq \lambda_c$ ($r_2 \simeq 0.1$)
to generate the correct charm mass. We have
$f_{11,12}$  $\sim 0$ (nonzero value of $r_2$ is
therefore allowed) to ensure small values of $h_{11,12}$,
and $h^\prime_{12}$ generates the down-quark mass and Cabibbo angle
$\theta_{c}$
with $m_d/m_s \simeq \sin^2\theta_c$.
The up-quark Yukawa coupling is generated as $\lambda_u \sim (r_2
\lambda^3)^2$, and we also have a relation $m_d m_s m_b \simeq c_e^2\, m_e
m_\mu m_\tau$, where $c_e^2 \simeq 1$ in the preferred vacuum for the $\bf 120$ Higgs coupling,
Eq.(\ref{su5-5}). In the basis where $Y_u$ is diagonal, $\tilde
A_{hh}$ in this texture is not completely zero but can become much smaller than
$10^{-8}$.

We comment that we also need to examine the contribution of the
 other components e.g. $\tilde A_{ff,hf,fh,h'f,h'h,\cdots}$ 
after we  suppress the  $\tilde A_{hh}$.
However,
their contributions are associated with the
colored Higgs mixing angles, which are inputs and can be suppressed by our choice of
the vacuum expectation values and the Higgs couplings.
According to our numerical studies, some of the mixing angles  must be
about a few percent in the case of $\tan\beta \sim 50$ to suppress the decay.
However, the mixing angles can become larger as $\tan\beta$ becomes smaller.
Details  will be given in a future paper\cite{dmm}.

Without the  particular choice of texture, as mentioned above, the
proton decay can not be suppressed naturally in these
models unless the $\tan\beta$ is very small.

The proton life time for $p\rightarrow K\bar\nu$ for this choice of texture
can be larger than the current experimental bound, $\tau_p
\agt 2\cdot 10^{33}$ years for any $\tan\beta$ (using the lightest
colored Higgsino mass to be  $2\cdot 10^{16}$ GeV and squark mass
scale  around 1 TeV). All other nucleon decay modes are
suppressed as well. In our calculation, we use long- and
short-distance renormalization factor, $A_L=1.43$ and typically
$A_S=1.8$ similar to Ref.\cite{goran}.

 \begin{figure}[t]
 \includegraphics[width=8cm]{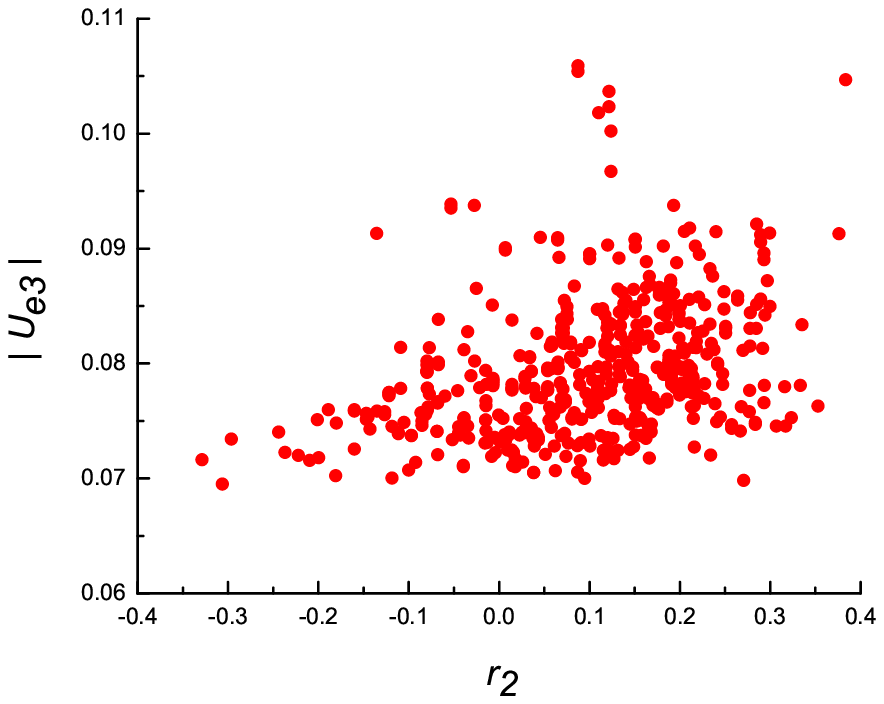}%
 \caption{\label{ue3} $|U_{e3}|$ is plotted as a function of $r_2$.}
 \includegraphics[width=8cm]{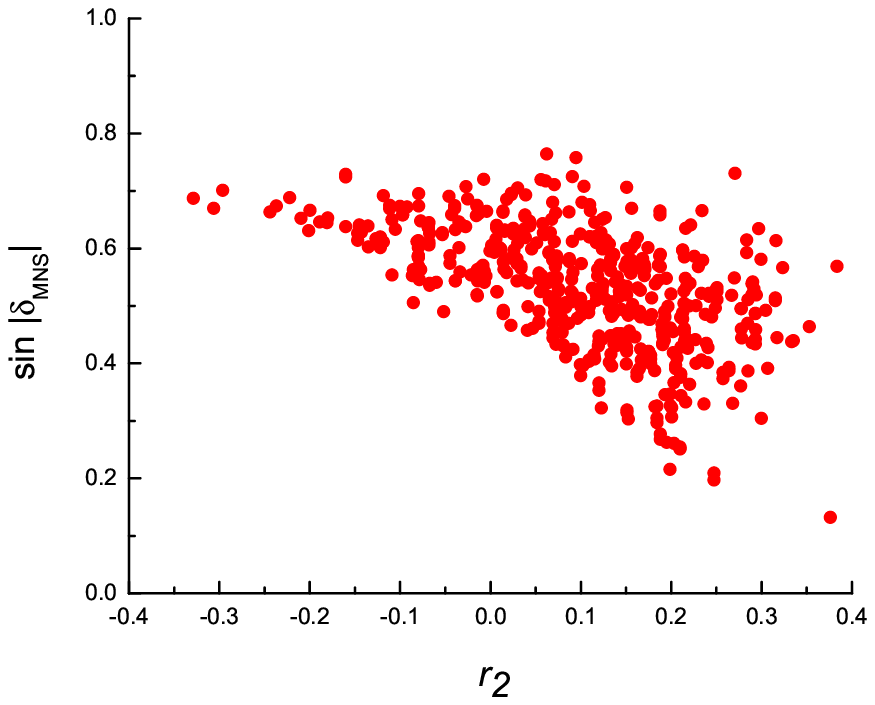}%
 \caption{\label{mns} $\sin|\delta_{MNS}|$ is plotted as a function of $r_2$.}
 \end{figure}

Given the above texture for Yukawa couplings, we find that $U_{e3}$ is
restricted to a range. In figure \ref{ue3}, we plot $U_{e3}$ as a
function of $r_2$
with $f_{11,12} \rightarrow 0$ (as required in the most preferred texture)
and the value at $|r_2|\simeq 0.1$ is most important. We find that
$U_{e3}$ is between 0.07 to 0.09.
For this fit, $\sin^22\theta_{23}$ is  maximal and $\tan^2\theta_{\rm
solar}\sim 0.4$.
The prediction for $\sin|\delta_{MNS}|$  lies between 0.3 to 0.7 for
$|r_2|\simeq0.1$ as shown in figure \ref{mns}. The Yukawa matrices are
assumed to be hermitian in order
to keep the model free from SUSY CP problem.
%

The presence of  $h'$ is a necessity  to suppress proton decay (suppress $\tilde
A_{hh}$) and fit the fermion masses.
This $h'$ also helps to explain CKM CP violation\cite{Dutta:2004wv}.

Finally, we show how the above proton decay suppression arises
by an adjustment among different VEVs 
(or symmetry breaking parameters).
 In the above, we
need to have $r_3 \simeq 0$ in the Eq.(\ref{Y_u}). Since this is
satisfied in Eq.(\ref{su5-5}) for the SU(5) condition, it may be a
hint that we stay close to the SU(5) symmetric vacuum. Secondly,
we need $r_2 \simeq 0$ while suppressing $\overline{\bf 126}$
colored Higgs contribution to proton decay, namely only $U_{14}$
is enhanced while other mixings for the sub-multiplets in
$\overline{\bf 126}$ are small,
\begin{equation}U_{14} \gg V_{15},\, X_{14}, \,Y_{15}.
\,\, 
\label{constraint}\end{equation}
%
In order to explain how these conditions may be satisfied in the
minimal SO(10) model, we denote the VEVs of the submultiplets in $\bf 210$
multiplet that break
SO(10) symmetry as follows: 
$\Phi_1 : ({\bf 1,1,1})$, $\Phi_2 : ({\bf 15,1,1})$, $\Phi_3 : ({\bf 15,1,3})$
(where numbers in the parenthesis denote $G_{422}$ quantum numbers).
The VEVs $\Phi_i$ are around the unification scale of three gauge couplings.
Recall
that in the SU(5) symmetric vacua\cite{Fukuyama:2004ps,Aulakh:2003kg}, the
$\Phi_i$'s satisfy the following relation:
$\sqrt6\Phi_1=\sqrt2\Phi_2=\Phi_3$
(We have used the same
normalization  as in the Ref.\cite{Fukuyama:2004ps}).
They lead to
 the SU(5) relations (\ref{su5-1}-\ref{su5-4}).
Now we perturb the Higgs potential with a small coupling, $\lambda_2 H
\Delta \Phi$.
We obtain $r_3 \propto \lambda_1(\sqrt6\Phi_1-\Phi_3)$
(where $\lambda_1$ is
associated with $\lambda_1HD\Phi$ term).
If $\sqrt6\Phi_1=\Phi_3$, we have $r_3\simeq 0$.
%
We obtain
the Higgs mixings
\begin{eqnarray}
U_{14} &\!\!\simeq&\!\! - 6\sqrt5\, \lambda_2/\eta
 \frac{\sqrt2 \Phi_2 -\Phi_3}{\sqrt6 \Phi_1 + \sqrt2 \Phi_2+ 8 \Phi_3} +
\cdots,\\
X_{14} &\!\!\simeq&\!\! - 2 \sqrt{15}\, \lambda_2/\eta
 \frac{\sqrt6 \Phi_1 - \sqrt2 \Phi_2}{\sqrt6 \Phi_1 + 3\sqrt2 \Phi_2+ 6
\Phi_3} + \cdots,\ \ 
\end{eqnarray}
where $\eta$ is a coupling of $\Phi\Delta\overline\Delta$ term. We also
have similar terms for $V_{15}$,
$Y_{15}$ and $Y_{16}$. All these terms have different denominators.
All the Higgs mixing angles tend to zero in the limit $\lambda_2\rightarrow0$.
However, suppose that $\sqrt6 \Phi_1 + \sqrt2 \Phi_2+ 8 \Phi_3 \sim 0$
is satisfied,
only $U_{14}$ can be of finite value, all other mixing angles are zero and
Eq.(\ref{constraint}) is satisfied.
On such a vacuum, the proper strange quark and the muon masses can be
realized by an
enhancement of $U_{14}$,
and Yukawa coupling of up-type quark, $Y_u$, is almost proportional to $h$.
This is just an example, and in our detailed quantitative work, we keep
all other terms in the Higgs potential
and we satisfy Eqs.(\ref{su5-5},\ref{constraint}) to suppress the proton
decay rate.
\section{Conclusion}
In conclusion, we have analyzed the fermion masses and dimension
five $\Delta B=1$ operators in the minimal SO(10) model with {\bf
10}, $\overline{\bf 126}$ and {\bf 120} Higgs fields coupling to
matter.
We show that by a choice of suitable textures, one can not
only get correct fermion masses and mixings but also suppress the
contributions to proton decay from both the $LLLL$ and $RRRR$
operators for the entire range allowed $\tan\beta$ parameter of
MSSM.
This choice of textures requires a suitable SO(10) breaking vacuum condition
which is close to SU(5) invariant vacua.
In the most favorable region of parameter space $U_{e3}$ is
predicted to be 0.07 to 0.09 and $\sin|\delta_{MNS}|$  to be
0.3 to 0.7, which can be used to test the model.
\section*{Acknowledgments}
This work of B.D. and Y.M. is supported by
the Natural Sciences and Engineering Research Council of Canada and
the work of R. N. M. is supported by the National Science Foundation
Grant No. PHY-0354401.


\end{document}